\documentclass[lettersize,journal]{IEEEtran}
\usepackage{amsmath,amsfonts}
\usepackage{algorithmic}
\usepackage{algorithm}
\usepackage{array}
\usepackage[caption=false,font=normalsize,labelfont=sf,textfont=sf]{subfig}
\usepackage{textcomp}
\usepackage{stfloats}
\usepackage{url}
\usepackage{verbatim}
\usepackage{graphicx}
\usepackage{cite}
\begin{document}

\title{The high/low frequency balance drives \\ the perception of noisy vibrations}

\author{Corentin Bernard, Etienne Thoret, Nicolas Huloux 
       and  S\o lvi~Ystad

\thanks{
Manuscript created October, 2023;

Corentin Bernard, Etienne Thoret and  S\o lvi~Ystad are with Aix Marseille Univ, CNRS, PRISM, Marseille, France.
E-mail : bernard@prism.cnrs.fr

Corentin Bernard and Nicolas Huloux are with MIRA, Aflokkat, Ajaccio, France.

Etienne Thoret is with Institute of Language Communication and the Brain, Aix-Marseille Univ, UMR7020 Laboratoire d'Informatique et Systèmes, UMR7289 Institut de Neurosciences de la Timone, CNRS, Marseille, France.

Nicolas Huloux is with Univ. Bordeaux, ESTIA-Institute of Technology, EstiaR, F-64210 Bidart, France.
}

\thanks{Etienne Thoret \& S{\o}lvi Ystad equally contributed to the supervision of this work. This work was supported by France Relance.
Author ET was supported by grants ANR-16-CONV-0002 (ILCB), ANR-11-LABX-0036 (BLRI) and the Excellence Initiative of Aix-Marseille University (A*MIDEX).}}


\markboth{Journal of \LaTeX\ Class Files,~Vol.~.., No.~.., October~2023}%
{Bernard \MakeLowercase{\textit{et al.}}: The high/low frequency balance drives the perception of noisy vibrations}


\maketitle

\begin{abstract} 

Noisy vibrotactile signals transmitted during tactile explorations of an object provide precious information on the nature of its surface. Linking the properties of such vibrotactile signals to the way they are interpreted by the haptic sensory system remains challenging.
In this study, we investigated humans’ perception of noisy, stationary vibrations recorded during exploration of textures and reproduced using a vibrotactile actuator.
Since intensity is a well-established essential perceptual attribute, an intensity equalization was first conducted, providing a model for its estimation.
The equalized stimuli were further used to identify the most salient spectral features in a second experiment using dissimilarity estimations between pairs of vibrations.
Based on dimensionally reduced spectral representations, linear models of dissimilarity prediction showed that the balance between low and high frequencies was the most important cue.
Formal validation of this result was achieved through a Mushra experiment, where participants assessed the fidelity of resynthesized vibrations with various distorted frequency balances. These findings offer valuable insights into human vibrotactile perception and establish a computational framework for analyzing vibrations as humans do. Moreover, they pave the way for signal synthesis and compression based on sparse representations, holding significance for applications involving complex vibratory feedback.

\end{abstract}


\section{Introduction}

\IEEEPARstart{S}{imple} vibrations have become a standard mean to convey information in our smartphones and game controllers. Recent technological advancements have expanded the frequency bandwidth of actuators, enabling software designers to create more precise vibrations. Additionally, the miniaturization of these actuators enables their integration into a wide range of human-machine interfaces, including wearables and vibrotactile touchscreens. These vibrations are primarily used to provide users with additional information, enhancing usability and accessibility. Moreover, finely-tuned vibrations have been employed to improve immersion in virtual gaming environments ~\cite{kim2006vibrotactile, singhal2021juicy} and facilitate remote social interaction~\cite{israr2018towards, lee2017exploring}.

Up to this point, two distinct approaches have been devised for generating vibrations. The first approach involves crafting signals using a simple model based on a limited set of parameters, often relying on a single sine wave with varying frequency and envelope~\cite{seifi2015vibviz}.
This method offers the advantage of being sparse while enabling the creation of a diverse range of stimuli. However, it falls short in accurately reproducing natural vibrations with a complex spectral profile. 
The second approach is based on the capture and faithful playback of vibrations induced by friction during texture exploration~\cite{culbertson2014modeling}.
This approach can achieve a high level of realism but is much more demanding in terms of amount of data to be restituted. 
The need to reduce data storage size has become increasingly crucial as new technologies emerge, incorporating a multitude of channels, such as multitouch haptic surfaces~\cite{pantera2021lotusbraille}, wearable devices~\cite{richards2022designing} or controllers  \cite{richard2023multivibes} equipped with multiple actuators.

On the other hand, the human body is consistently exposed to vibrations generated by the friction between the skin and external surfaces, especially when manipulating objects with the fingers. 
This influx of information is initially processed by mechanoreceptors, which transform mechanical vibrations into electrical spike patterns~\cite{hollins2007coding, weber2013spatial, manfredi2014natural}, 
The efficient coding hypothesis~\cite{barlow1961possible} postulates that the sensory system encodes incoming information as efficiently as possible, eliminating redundancies to minimize the number of spikes and reduce neural activity.
In touch, prior research has suggested potential reduction in dimensionality when encoding skin deformations~\cite{hayward2011there}, such as for detecting of object slippage~\cite{willemet2022efficient}.


In this paper, we investigated human vibrotactile perception to identify perceptually relevant signal features that could further be used to simplify the vibrotactile signal. Our hypothesis was that by extracting these features from the incoming vibrations, we could reduce the signal to what is strictly perceptible.

Through psychophysical experiments, we identified spectral structures that perceptually stand out within noisy vibrations. 
We demonstrated that any signal can be projected onto this perception-driven basis and then reconstructed with this reduced information.
In addition to the interesting synthesis perspective of this approach, it can also be employed for achieving sparse representations for vibrotactile signal compression.

The paper is divided into three sections, each corresponding to a distinct experiment dedicated to the perception of noisy stationary vibrations, which were here recordings of friction-induced vibrations that present temporal homogeneity. 
The first experiment investigates the perception of the intensity of these vibrations.
It provides a relationship between signal power and perceived intensity, as well as an intensity equalization of the stimuli, hereby preparing for the next experiments.
The second experiment dvelves into the perception of other attributes by posing a simple question: which spectral attributes enable distinguishing sensations between two vibrations? 
The results underscored the essential role played by one fundamental element in vibrotactile perception: the balance between high and low frequencies. 
Lastly, the third experiment not only reaffirms these previous findings but also showcases their potential in terms of vibrotactile synthesis and compression.

\section{Background}

\subsection{Intensity and frequency perception}

The literature on vibrotactile perception offers a plethora of studies that have focused on two key characteristics of simple vibrations: intensity and frequency. 
Intensity is known as the principal perceptual attribute in vibrotactile perception, playing a pivotal role in discriminating between tactile stimuli. This holds true for both simple synthesized vibrations~\cite{hwang2010perceptual} and for friction induced vibration recorded from various textures~\cite{felicetti2023tactile}.
Perceived intensity is directly associated with the signal's amplitude. Remarkably, humans can discern subtle differences in intensity regardless of the frequency, with discrimination thresholds ranging from 11\% to 30\% depending on the study (see \cite{craig1972difference} for a review).
Moreover, it is noteworthy that perceived intensity is also influenced by the signal's frequency. Amplitude detection thresholds follow a U-shaped curve related to frequency, 
similar to curves of equal intensity~\cite{verrillo1969sensation}.
Vibrations of equal amplitude are perceived as more intense at frequencies around 200~Hz.
Models have been developed to predict the intensity of a single sine wave based on its amplitude and frequency~\cite{wang2008constructing}, and also for signals comprising a sum of two sine waves~\cite{yoo2022perceived}. 
Furthermore, humans possess the ability to distinguish frequency differences when they deviate by more than 20\%. The just-noticeable difference typically falls between 17\% and 21 \% across various studies, following Weber's Law~\cite{pongrac2006vibrotactile,pongrac2008vibrotactile}.

The literature also sheds lights on the frequency selectivity of vibrations composed of two distinct frequencies. When these two frequencies are sufficiently distant, separated by more than 100 Hz, they give rise to a unique percept.
In such cases, we are unable to discern the two individual vibrations distinctly; instead, we perceive a vibration that falls somewhere between the initial frequencies, with the perception being influenced by the amplitude of each frequency component~\cite{hwang2017perceptual,bernard2018harmonious,friesen2018single}. 
Conversely, when the two frequencies are closely spaced ($<$100~Hz), interference patterns emerge, resulting in what is known as "beating". In this scenario, the frequency of the beating, or the frequency of the amplitude modulation, takes on a more prominent role as a perceptual attribute compared to the initial frequencies~\cite{park2011perceptual,bernard2022tactile}, inducing a sensation of rhythm~\cite{bernard2022rhythm}.

From a physiological perspective, perceived intensity is primarily mediated in the somatosensory periphery by the total population of nerve fibers that are activated~\cite{bensmaia2008tactile}, whereas the frequency coding relies on temporal spiking patterns~\cite{birznieks2017spike}.

\subsection{Vibration encoding and compression}

Signal encoding is a crucial issue for vibration rendering. 
Since humans are sensitive to vibrations up to 1000~Hz~\cite{verrillo1969sensation}, temporal signals are usually encoded with a sample rate of about 2000~Hz. An 8 bits quantization of the signal has been shown to be sufficient for preserving the perceived quality~\cite{consigny2022perceptual}.

Previous studies have also demonstrated that classical mathematical transformations, such as the Fourier transform or the Gabor transform, were relevant to model the encoding of vibration signals in accordance to human perception~\cite{wiertlewski2011spatial,meyer2015modeling,toide2023sufficient}.

In the literature, many contributions have studied the compression of vibrotactile signals in order to develop the best methods to reduce the file size without altering its tactile quality.
The main idea is to remove information that is not perceived from the signal.
Previous works focused at removing frequency components that are below the detection threshold~\cite{okamoto2012lossy,hassen2020pvc}. 
Based on frequency masking~\cite{gescheider1982prediction}, other approaches \cite{chaudhari2014perceptual,noll2021vc, nockenberg2023mvibcode} consist in removing the low-amplitude frequencies that are imperceptible due to their vicinity with a high-amplitude frequency.
A measure of vibration similarity, based on spectral and temporal similarities~\cite{hassen2019subjective}, has been developed to assess the compression quality and compare compression methods~\cite{muschter2021perceptual}.

\section{Material and methods}

This section outlines the setup utilized in the three experiments presented in this paper. The experiments were approved by the Ethical Committee of Aix-Marseille University.

\subsection{Vibrotactile stimuli}

Recordings of friction-induced vibrations from Kirsch et al.'s database~\cite{kirsch2018low} were used to provide vibrotactile stimuli. These signals were recorded during exploration of textures
with a specific tool equipped with an accelerometer.
We selected 18 signals corresponding to 9 materials (rubber, polyester pad, foam, felt, cork, bamboo, baltic brown (granite), anti-vib pad (recycled rubber) and aluminium grid) explored by 2 different probes (round and spiked) for a medium scanning speed condition.
To use stationary signals, one second of each recording during which the scanning speed was roughly constant between 100 and 120 mm/s was kept.
The sampling rate was set to 2800~Hz.

\subsection{Vibration rendering}

\begin{figure}[h!]
\centering
\includegraphics[]{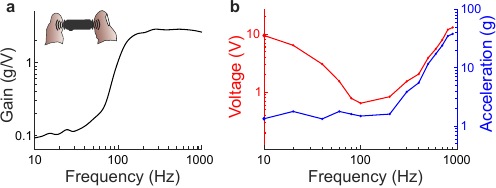}
\caption{
\textbf{a.} Frequency response of the actuator held by two fingers, as in the experimental conditions.
\textbf{b.} Iso-intensity vibration curve rendered by the actuator and felt with two fingers. 
The curves display the peak-to-peak voltage of the input signal (in red) and the peak-to-peak acceleration (in blue) corresponding to sine-waves at various frequencies that are perceived with the same intensity.
}
\label{fig:actuator}
\end{figure}

Tactile stimuli were presented through an Actronika (Paris, France) HapCoil-One vibrotactile actuator (dimensions, $11.5 \times 12 \times 37.7~\mathrm{mm^3}$; frequency bandwidth,
10–1000 Hz; resonant frequency with no load, 65 Hz). The actuator was powered by a Pioneer A-209R audio amplifier. The participants were asked to grab the vibrotactile actuator between the thumb and the index finger of their left hand to feel the vibrations.
Fig.~\ref{fig:actuator}.a presents the frequency response of the actuator while being held with two fingers.
In the following, vibrations will be described  either by their input voltage when considering the source signal (for intensity and synthesis), or by their acceleration (signal filtered by the frequency response) when investigating human perception.

The actuator was also characterized \textit{via} its iso-intensity curves presented in Fig.~\ref{fig:actuator}.b. The intensity judgments were gathered in a previous experiment involving five participants, with the  method of direct intensity matching with sine-waves as in~\cite{verrillo1969sensation}.
The curves show in which frequency band participants are most sensitive to the actuator, in terms of input voltage (proportional to displacement) or acceleration.

During all experiments, participants wore noise-canceling headphones playing pink noise to mask sounds produced by the actuator and avoid any auditory bias.

\section{Intensity perception}

As detailed previously, intensity is well known as the primary perceptual attribute of vibration. 
However, existing models of intensity perception in the literature have primarily focused on simple vibrations with one or two sine waves. To date, there are no intensity models specifically designed for noisy vibrations.
 
Consequently, a preliminary experiment was conducted to equalize the intensity of the stimuli, so that the contribution of other attributes could  be studied regardless of the intensity.
Besides, the participants' judgments were examined to develop a model of perceived intensity for noisy vibrations.

\subsection{Protocol}

A mathematical amplitude normalization was first conducted by equalizing the standard deviation of the 18 vibration signals.
Perceptual intensity equalizations were then performed by the participants who were asked to adjust the gain of each stimulus to match the intensity of a reference stimulus: a white noise with normalized standard deviation amplitude.
For the 18 stimuli, the gain could be adjusted between $0.2$ and $5$, while the reference was assigned a gain of 1.

\subsection{Participants}

The intensity equalization was performed by 10 participants, 2 females and 8 males, from 23 to 55 years old (mean=29.7), 1 left-handed and 9 right-handed.

\subsection{Results}

Participants responses were coherent, as shown by high pairwise correlations between participants on the selected gains (Pearson's r: M=0.82, SD=0.08, df=16 for the 45 correlations, all significant p$<$0.05).

For each stimulus, the average gain was computed as the geometric mean of all participants judgments. The mean gains were then applied to the signals in order to obtain the 18 iso-intense vibrotactile stimuli for the second experiment.
The high correlations (Pearson's r: M=0.91, SD=0.05, df=16 for each of the 10 correlations, all significant p$<$0.05) between the mean gains and the participants individual gains shows that the mean gains faithfully represent the participants judgments.

\begin{figure}[h!]
\centering
\includegraphics[]{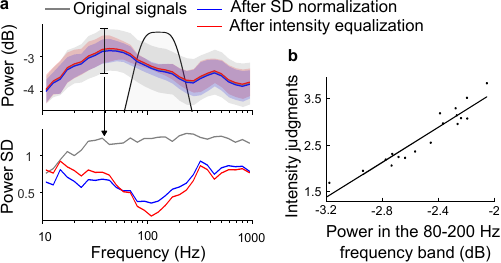}
\caption{
\textbf{a.} Results of the equalization of the perceived intensity for the 18 vibrations.  
The original signals from the Kirsch et al. database~\cite{kirsch2018low} are displayed in grey. These signals were equalized in standard deviation prior to the experiment (in blue). The signals equalized in intensity by the participants are shown in red.
On the top, the mean power spectra of the signals are displayed in continuous line and the standard deviation in shaded area.
On the bottom, only the standard deviation is displayed for comparison.
\textbf{b.} Linear modeling of intensity prediction ($R^2=0.90$).
The dots represent the mean intensity judgments for each stimulus with respect to its power in the 80-200~Hz frequency range, calculated with the filter showed on the top-left (not to scale).
}
\label{fig:intensityEqualization}
\end{figure}

Fig.~\ref{fig:intensityEqualization}.a presents the result of the intensity equalization on the power spectra of the stimuli. After equalization, the vibration signals show less variability in the 80-200~Hz frequency band. This means that, when asked to equalize the intensity, the participants tended to equalize the power of the signals in this specific frequency band, regardless of the power in higher or lower frequency bands.

\subsection{Model of intensity perception}

Based on these results, we constructed a model to predict the perceived intensity of noisy vibrations from their signal. The intensity judgments were obtained as the inverse of the selected gains.
For each signal, its power in the frequency band of interest (80-200~Hz) was computed using a $2^{nd}$ order Butterworth bandpass filter. 
The linear regression curve ($R^2=0.90$, df=16), displayed in Fig.~\ref{fig:intensityEqualization}.b, demonstrates that we can accurately model the intensity judgments from the power in that frequency band.
The goodness of fit dropped when increasing the frequency bandwidth, to $R^2=0.07$ for the whole frequency band (10 to 1000~Hz).
This means that, for noisy vibrations, the perceived intensity is
directly determined by the power of the signal in the 80 to 200~Hz frequency band, and weakly influenced by the power of higher and lower frequencies.
This frequency band matches the optimal sensitivity band outlined in Fig.~\ref{fig:actuator}.b, which corresponds to the combination of human tactile sensitivity and the actuator frequency response.

\section{Spectral content perception}

The main experiment focuses on the perception of the spectral content.
With the stimuli now having equal intensity, it becomes more straightforward to explore how other attributes influence vibrotactile perception.
We used a protocol based on dissimilarity ratings, wherein participants were asked to rate the (di-)similarities between pairs of stimuli. This technique is a gold standard in the field of auditory perception to investigate the perception of sound timbre for musical instruments~\cite{mcadams1995perceptual, grey1977multidimensional,barthet2010clarinet}, see \cite{thoret2021learning} for a recent meta-analysis.
It enables to model perception within a reduced dimensional space and to identify the main perceptual dimension within this space. Through further analysis, links can be established between these dimensions and signal features, ultimately unveiling the association between signal and perception.

This approach has also demonstrated its potential in the study of touch. For textures, perceptual differences can be predicted using the comprehensive data collected during their tactile exploration, encompassing forces, vibrations, and velocity ~\cite{richardson2022learning}.
A recent study~\cite{lim2023can} proposed a physiology-based model to predict the dissimilarities in vibrations designed with varying amplitudes, frequencies and modulation rates.

Here, as proposed in~\cite{thoret2021learning}, we first gathered dissimilarity ratings from human participants and then trained a linear model, fully interpretable, to predict the obtained dissimilarities based on the spectral representations of the stimuli.
We assume that once the proposed model is trained to mimic human responses, we can interpret its behavior to gain insight on human perception.

\subsection{Participants}

The dissimilarity rating experiment was performed by 18 participants, 3 females and 15 males, from 21 to 55 years old (mean=28), 1 left-handed and 17 right-handed.

\subsection{Protocol}

Participants had to evaluate the perceptual differences of the vibrations through pairwise comparisons. The 18 stimuli were presented against each other, resulting in 153 pairs. The stimuli were also compared with themselves to provide a baseline for the most similar ratings, leading to 171 pairs in total, presented in random order.
At each trial, participants felt the two vibrations successively. The vibrations lasted for 1~second, and were separated by  500~ms of silence.
Each sequence was played once.
Participants were asked to judge the perceived dissimilarities between the pairs and to report their ratings on a scale from "very similar" to "very dissimilar" thanks to slider on a visual interface.
The interface converted the slider position into a dissimilarity rating between 0 and 1 (0=very similar, 1=very dissimilar).
Prior to the experiment, a familiarization session was  carried out with 30 random pairs. It enabled the participants to  familiarize themselves with the task and the stimuli, and to create their own internal rating scale.
The experiment lasted for about 45~min.

\subsection{Results}

Regarding dissimilarity ratings, participants were less consensual than for the intensity ratings, as revealed by the pairwise correlation scores between participants (Pearson's r: M=0.51, SD=0.12, df=169 for the 162 correlations, all significant p<0.05). Yet, as the evaluation strategies were coherent across participants, we averaged their dissimilarity scores.
The correlations between the average and the participants' scores (Pearson's r: M=0.73, SD=0.09, df=169 for the 18 correlations, all significant p<0.05) showed that the mean dissimilarity scores well reflect participants' judgments. The following analysis therefore focuses on the between-participants mean dissimilarity scores.

For pairs with the same stimulus, the dissimilarity scores were low (M=0.12, SD=0.07), but still higher than 0. This means that the task was quite difficult and the intensity-equalized vibrations felt already pretty similar.
It also means that vibrations with a dissimilarity score below 0.12 could be considered as perceptually identical.

\subsection{Predicting dissimilarities from spectral representation}

\begin{figure}[h!]
\centering
\includegraphics[]{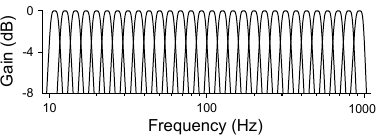}
\caption{
Frequency response of the filter bank used to compute the power by frequency band of the vibration signals. The 30 filters are logarithmically spaced following Weber's law to match human perception.
}
\label{fig:filterbank}
\end{figure}

\begin{figure*}[h!]
\centering
\includegraphics[]{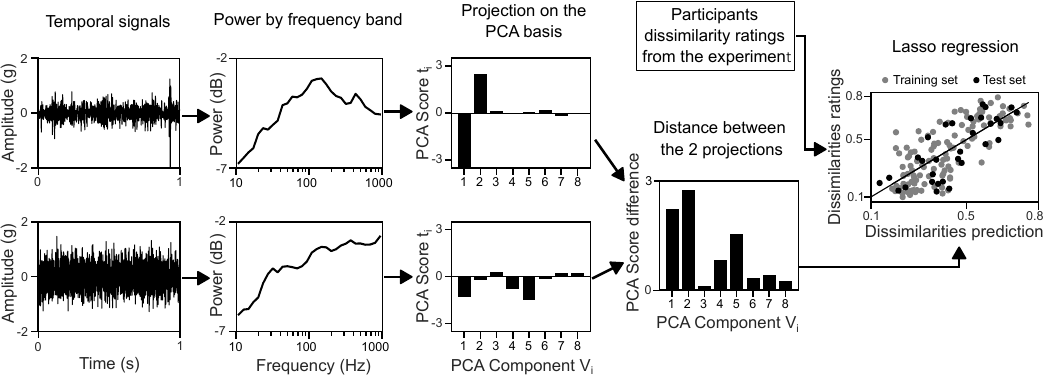}
\caption{
Principle of the dissimilarity prediction model, illustrated here to compare two vibrations.
The spectra are projected on a basis of spectral patterns given by the PCA. Then, the distance between these two representations is computed.
Weights are associated to each component and are optimized so that the mathematical distances fit the perceptual dissimilarities as closely as possible.
}
\label{fig:methodology}
\end{figure*}

Since the stimuli were considered as stationary, i.e. they were felt to be relatively constant during the 1 second presentation time, we chose to represent the signals by their power spectra to focus on the spectral properties.
The power spectrum was computed as the power by frequency band using a filter bank with 30 second-order bandpass-filters logarithmically spaced between 10 and 1000~Hz, as shown in Fig.~\ref{fig:filterbank}.
This representation of the signal power spectrum offers the advantage of being both sparse and close to perception.
It covers the human vibration sensitivity range and the filter distribution follows  Weber's law of vibrotactile frequency perception, with a Just-Noticeable frequency Difference (JND) of 17\%  \cite{pongrac2008vibrotactile} between two filters: 
$N=log(f_{max}/f_{min})/log(1+JND) \approx 30 $ filters.

Continuous spectral patterns were observed, indicating that the signal representation could be further reduced.
Therefore, a Principal Component Analysis (PCA) was conducted to provide a data-driven dimensionality reduction. The PCA was trained on the power by frequency band of all the signals in the Kirch et al. database \cite{kirsch2018low} (composed of 281 signals). 
We found that the first 8 dimensions of the PCA were sufficient to capture 95\% of the variance of stimuli in the database.
This enabled us to decompose the experiment's 18 stimuli into the most important spectral patterns of friction-induced vibrations presented in Fig.~\ref{fig:PCA}.a.
We can already note that the first component $V_1$ encodes a balance between high and low frequencies, and other components encode spectral patterns of increasing complexity.

Based on the spectral patterns representations of the vibration, we constructed a model to predict the participants' dissimilarity ratings from the experiment.

The methodology used to compare two vibrotactile signals $x$ and $y$ is presented in Fig.~\ref{fig:methodology}. The algorithm takes the two temporal signals as input.
Firstly, the power spectra are measured by computing the power by frequency band $P_x$ and $P_y$ ($dim=30$, using the filter bank described previously). Then, the power spectra  are projected on the PCA basis $V$ ($dim=8\times30$) to obtain the PCA scores $T_x$ and $T_y$ ($dim=8$):

 \begin{equation}
     T_x=V P_x  \mathrm{~~and~~}  T_y=V P_y
 \end{equation}

Next, the local distance $d$ in each PCA dimension $j$ is calculated as the absolute difference between the two PCA scores.

\begin{equation}
	 d_j(x,y)=\sqrt{(T_{xj} - T_{yj} )^2  }
\end{equation}

Finally, the global dissimilarity $D$ between the two vibrations is computed as a weighted sum of the local distances:

\begin{equation}
	D(x,y)=\sum_{j=1:8} w_j d_j(x,y) 
\end{equation}

The weights $w_j$ were optimized so that the dissimilarity prediction best matched the participants' dissimilarity ratings. It was performed by a Lasso regression, a multiple regression model with regularization that performs variable selection to facilitate the interpretation of the results.

To train the model, while the best amount of penalization was chosen by cross validation (10-fold), the weights were optimized on a training set of 122 dissimilarity ratings (80\% of the data). The prediction performance of the model was then evaluated on a test set of 23 dissimilarity ratings (20\% of the data).

For the prediction, the model exhibited a coefficient of determination $R^2=0.54$ (mean of 100 repetitions with random assignment to the training and test sets, SD=0.11). The prediction score is not as high as for the intensity model, but the participants were also less coherent. Still, more than half of the variance is explained.
\subsection{Interpretation of the dissimilarity model}

\begin{figure}[h!]
\centering
\includegraphics[]{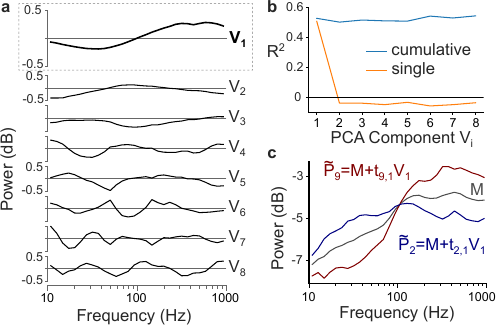}
\caption{
\textbf{a.} Representation of the 8 PCA components $V_{i}$ to highlight spectral patterns.
\textbf{b.} Contribution of each PCA component to the prediction quality. The red curves show the evolution of the coefficient of determination when the prediction is performed using the projection of the signals on the $i^{\rm th}$ component only.
The blue curves show the evolution of the coefficient of determination when the prediction is performed with projection on the basis made of the $i$ first components.
\textbf{c.} 
Example of signal spectra (for stimuli $s=2$ and $s=9$) reconstructed only from the score on the first PCA component $t_{s,1}$. M is the averaged spectrum of the database.
}
\label{fig:PCA}
\end{figure}

The linearity of the proposed model offers the advantage of being easily interpretable.
We can assess the importance of a PCA component by examining the impact of its withdrawal on the prediction quality $R^2$.

Fig.~\ref{fig:PCA}.b shows that the prediction is only based on the projection of the stimuli on the first PCA component $V_1$. Indeed, the prediction scores $R^2$ are below 0 when other components are considered alone (red curve) and $R^2$ does not increase as more PCA components are included (blue curve).
This means that, to mimic human judgments, the model only needs information about the first PCA component, i.e. the balance between high and low frequencies.

\begin{figure}[h!]
\centering
\includegraphics[]{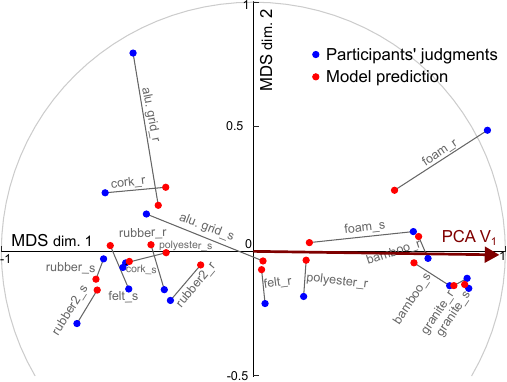}
\caption{
Two-dimensional perceptual space resulting from the MDS. 
The position of the 18 stimuli according to the participant's judgments is plotted in blue. Their position resulting from a second MDS based on the dissimilarities predicted by the model is plotted inr red.
The error between the two is revealed by the grey lines.
The arrow represents the projection of the first PCA component $V_1$ in the MDS space and the unite circle is plotted for comparison.
The materials underlying the friction-induced vibrations are indicated, as is the type of probe: round (r) or spiked (s).
}
\label{fig:MDS}
\end{figure}

A classical multidimensional scaling analysis (MDS) was conducted to place each stimulus in a 2-dimensional space so that the distances between stimuli, regarded here as perceptual dissimilarities, would show up as clearly as possible.
The blue points in Fig.~\ref{fig:MDS} show the result of the MDS performed on the participant's mean dissimilarity ratings.
The projection of the first PCA component $V_1$ in this space demonstrates its high correlation (Pearson's $r=0.94$, df=16, p$<$0.01) with the first MDS dimension (and $r=-0.01$, df=16, p=0.95 with the second dimension).
In comparison, the other PCA components are much lesser correlated with the MDS dimensions (Pearson's $r \in [-0.4, 0.5]$ for all correlations, df=16, non-significant).
The first dimension of the MDS is known to reveal the main stimulus property rated by the participants. This analysis confirms the importance of the $V_1$ basis, the balance between the high and low frequencies, in vibrotactile perception.

An MDS analysis was also conducted on the dissimilarities given by the model. 
It was combined with a Procrustes superimposition, a combination of translations, rotations and uniform scaling, to match the two spaces and enable comparison. 
The positions resulting from predicted dissimilarities are displayed in blue in Fig.~\ref{fig:MDS}. The error distances, in grey, show that the model better predicts dissimilarities with certain stimuli.

Many participants spontaneously reported after the experiment that a few stimuli presented intensity temporal variations. Therefore, we computed a non-stationarity metric defined as the standard deviation of perceived intensity over time, using the previous model of intensity estimation by time windows of 100~ms, with 50~ms overlap.
This metrics appeared as well correlated with the second dimension of the MDS (Pearson's $r=0.86$, df=16, p$<0.01$).

It is also interesting to note that vibrations induced by friction on similar materials are located in the same zones of the perceptual space.

\section{Validation through analysis by synthesis}

The previous experiment highlighted the prominence of the balance between the high and low frequencies to feel dissimilarities between vibrations. 
We will now evaluate the validity of the previous insights by evaluating vibrations recreated by synthesis using this frequency balance.
This section also shows how this property of the human tactile sensory system can be leveraged to propose sparse vibration synthesis.

\subsection{Synthesis procedure}

\begin{figure}[h!]
\centering
\includegraphics[]{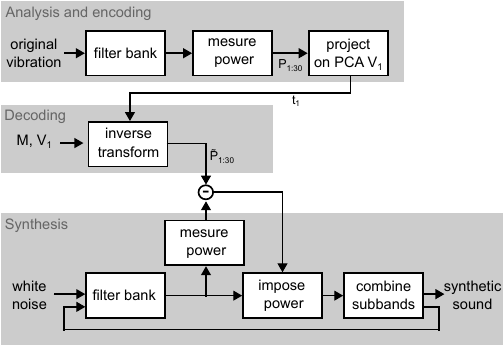}
\caption{Procedure of the analysis-synthesis.
The power by frequency band of the original vibration is projected on the axis $V_1$ (first component of the PCA on the whole vibration database).
The synthetic signal is based on noise, to which the powers of the frequency bands from the first PCA component are imposed. This algorithm was derived from~\cite{mcdermott2011sound}.
}\label{fig:synthesis}
\end{figure}

The proposed approach is based on analysis by synthesis, a well-known approach in audio processing, that consists in extracting parameters from a given natural sound to reconstruct it using algorithmic techniques~\cite{risset1999exploration, kronland2001sound}. 
Analyzing the structure of the algorithm with respect to the quality of synthesized sound has proven its potential for probing human perception.~\cite{mcdermott2011sound}.

The analysis-synthesis procedure is described in Fig.~\ref{fig:synthesis}. The original vibration, a stationary signal, is analyzed by filtering it with the logarithmic 30-filters bank. The power of each frequency band $P_{1:30}$ is then computed and projected on $V_1$ the first component of the PCA (the PCA that have been performed on the whole database~\cite{kirsch2018low}).
The input vibration is thus encoded as a unique scalar: the first principal component score $t_1$. 

The decoding part requires the average power by frequency band of the database $M$ and the first component of the PCA $V_1$. These features are the same for each vibration in the database. From the input $t_1$, the inverse transform is performed to obtain an approximation of the powers of the 30 frequency bands $\Tilde{P}_{1:30}$.
The Fig.~\ref{fig:PCA}.c shows examples of two signal spectra reconstructed by this method.

The synthesis procedure takes noise as an input, filters it, measures the power in each frequency band and applies the desired values $\Tilde{P}_{1:30}$. The output is obtained by combining the subbands, but this simple process does not produce exactly the desired signal. Therefore, an iteration procedure is conducted until the powers of the bands perfectly match  $\Tilde{P}_{1:30}$, and the algorithm outputs the synthesized signal.

\subsection{Protocol}

A third experiment was conducted to evaluate the quality of the sparse synthesis that we developed. The protocol was similar to a MUSHRA (Multiple Stimuli with Hidden Reference and Anchor).
At each step, the participants were instructed to rate and rank the similarities between a reference stimulus (the original vibration) and 5 stimuli:
\begin{itemize}
\item the synthesized version of the vibration $S_1$ using the proposed sparse algorithm (with the first component $V_1$ only)
\item a synthesized version of the vibration  $S_{1:8}$ with all the information (with the 8 components $V_{1:8}$)
\item a synthesized version of the vibration $S_{2:8}$ with all the information except the first component (with $V_{2:8}$)
\item the original vibration $O$ as control (the hidden reference) 
\item a synthesized vibration based on the mean power spectrum $M$, used as a control (anchor) and identical for all the stimuli.
\end{itemize}
The stimuli were 1~second vibrations that the participants could play as many times as they wanted by clicking on the corresponding buttons
The interface displayed 5 sliders (in random position) to rate the 5 versions of the stimuli on a scale from "very dissimilar" (0) to "very similar" (1). This task was performed 18 times for the 18 stimuli from the previous experiment, presented in random order.
The participants were asked to score the most similar stimulus to 1. Preliminary tests showed that it was preferable not to ask to score the most dissimilar stimuli to 0 (as in classic MUSHRA methods), since some of the original stimuli were very close to the anchor $M$.

\subsection{Participants}

The evaluation of the synthesized vibrations was performed by 15 participants, 6 females and 9 males, from 22 to 55 years old (mean=30), all right-handed.

\subsection{Results}

\begin{figure}[h!]
\centering
\includegraphics[width=8cm]{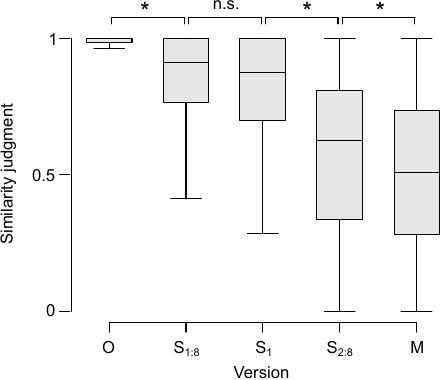}
\caption{Evaluation of the quality of the synthesized stimuli.
The 18 original vibrations were compared against themselves (O) and synthesized versions based on all the PCA dimensions ($S_{1:8}$), based only on the first dimension ($S_1$), based on all the dimensions except the first one (($S_{2:8}$) or based on the mean spectrum (M).
}
\label{fig:boxplot}
\end{figure}

The results of the comparison between the original vibrations and the synthesized versions are presented in Fig.~\ref{fig:boxplot}.

A two-ways repeated measures ANOVA, showed a significant effect ($\alpha=0.05$) of the version: F(14,4,17)=33, p=$1e^{-16}$) and Tuckey post-hoc tests are summarized in Fig.~\ref{fig:boxplot}. 
Most importantly, it showed that $S_1$ was significantly rated as more similar to the reference than  $S_{2:8}$ (p$<$0.01) but not less similar than $S_{1:8}$ (p$=0.76$).
A test of equivalence~\cite{wellek2010testing} showed that $S_1$ and $S_{1:8}$ were equivalent (at $\alpha=0.05$) in a $\pm 0.15$ interval on the 0 to 1 similarity scale, demonstrating that the first PCA component $V_1$ is necessary and sufficient to capture all the spectral information used by subjects to rate the dissimilarity between two textures.

However, the post-hoc tests highlighted significant differences between the synthesized versions and the original signals (p$<0.01$). They were mainly due to three stimuli (\textit{alu\_grid\_r}, \textit{alu\_grid\_s} and \textit{foam\_r}) that showed large discrepancies between $S_1$ and $O$.
This means that certain stimuli were not accurately synthesized by the the algorithm.
These stimuli were also the ones that were not well predicted by the model in Fig.~\ref{fig:MDS}. Moreover, the differences in similarity ratings between $O$ and $S_1$ for each stimulus were correlated (Pearson's $r=0.72$, df=16, p$<$0.01) with the prediction error of the model in the MDS space (grey lines in Fig.~\ref{fig:MDS}).
Also, we found that the differences in similarity ratings between $S_1$ and $M$ for each stimulus were correlated (Pearson's $r=0.63$, df=16, p$<0.01$) with the absolute values of their projection on the first dimension $|t_1|$. In other words, the more the frequency balance differs from the mean spectrum, the more the stimulus is perceived as dissimilar to $M$. This is an additional argument supporting the importance of the frequency balance in vibrotactile perception.

\section{Discussion}

The present work investigated the perception of stationary noisy vibrations, such as recordings of friction-induced vibrations with a constant exploration speed and aimed at unveiling their perceptually relevant signal structures.

\subsection{Intensity perception}
Firstly, the perception of intensity, the primary attribute of vibration, was investigated.
We discovered that, akin to simple sinusoidal vibrations, it is possible to gauge the perceived strength of a noisy vibration from its signal.
The results revealed that the intensity was directly proportional to the power of the frequency band to which the subjects were the most sensitive. The frequency band was between 80 and 200~Hz and corresponded to a combination of the human tactile sensory curve for sine waves~\cite{verrillo1969sensation} and the actuator's frequency response. The power in the higher and lower frequencies, to which we are less sensitive, had a negligible effect on intensity perception. This model can be easily adapted to other vibrating devices just by measuring their frequency response.
Equalizing the intensity of the vibration enabled us to remove this essential characteristic from subsequent experiments, enabling a detailed exploration of other attributes such as the spectral content.
\subsection{Spectral content perception}
The power spectra of the vibrations were computed thanks to a perception-based logarithmic filter bank. The principal component analysis of the spectrum of 281 recorded vibrations provided a data driven basis whose first axis encoded a balance between the low and high frequencies.
The projection of the vibrations solely onto this axis (a single scalar) was sufficient to build a model of dissimilarity prediction, and showed a high correlation with the first dimension of the MDS.
Moreover, vibrations that were re-synthesized using only this axis were indistinguishable from vibration re-synthesized using the whole spectral information.
These findings clearly demonstrate that, regarding human perception, the spectral content can be characterized by the balance between the low and high frequencies only.
This outcome joins previous works that showed that the spectrum of friction induced vibrations could be modelled by a $1/f^{\alpha}$ function~\cite{wiertlewski20111}. The parameter $\alpha$ also represents the balance between low and high frequencies and has been shown to correlate with perceptual judgements of texture categories~\cite{toscani2022database} and texture pleasantness~\cite{klocker2013physical}.
From a physiological perspective, the frequency balance could reflect the dual neural mechanism of flutter-vibration~\cite{talbot1968sense, mountcastle1967neural}. The balance could be the relative importance of the activation of FA I fibers (Meissner's corpuscles) in response to frequencies in the flutter range ($<60$~Hz) compared to the activation of FA II fibers (Pacinian corpuscles) in the vibration range ($>60$~Hz).

However, the model of dissimilarity estimation was not perfectly accurate, especially with some stimuli such as \textit{alu\_grid\_r}, \textit{alu\_grid\_s} and \textit{foam\_r} as shown in Fig.~\ref{fig:MDS}. Similarly, the synthesis algorithm failed to properly reproduce these stimuli.
We assume that these differences were due to the lack of stationnarity, since the friction recordings were not perfectly controlled. In particular, some participants reported that \textit{alu\_grid\_r}, \textit{alu\_grid\_s} were felt as not constant. 
The metric of non-stationarity suggests that the second dimension of the MDS may reflect temporal variations.
For vibrations that deviate from the assumption of temporal homogeneity, important information lies in the phase and is therefore not captured by the power by frequency band on which the model is based. 
This shows the limitations of the present analysis to stationary signals and further work will explore ways of including temporal variations (such as time windowing).

Another limitation of the analysis is its dependence on the data base. The vibrations have all been recorded with the same device and the same protocol and the three experiments have been performed with 18 stimuli only. However, the projection axis $V_1$ at the core of the analysis has been obtained by a PCA on 281 signals, presenting a wide variety of textures, exploration tools and exploration speeds. We are confident that the analysis procedure and the synthesis principle can be generalized to other vibration databases. 

\subsection{Sparse synthesis and compression}
The benefits of the study are considerable in terms of vibration synthesis and vibration compression.
It enables the creation of a sparse, perception-based synthesizer that can replicate a diverse range of texture-like vibrations using only two controls: intensity and frequency balance. 

Very high compression can also be attained through analysis-synthesis, by extracting intensity and frequency balance from the original signal and then using these parameters to synthesize a new vibration that closely resembles the original.
For example, in our case, the 18 signals recorded over 1 second at a sampling rate of 2800~Hz yield a total of $18 \times 2800 =50400$ data points.
The synthesis algorithm requires the power of the 30 frequency bands for the mean spectrum ($M$) and the first PCA component ($V_1$), and the projection scores ($t$) for each stimulus. This results in a total of $30+30+18\times 1=78$ data points for the encoded data, making up  0.15\% of the initial data, without considering quantization.
Since we are considering only stationary signals, the signal duration does not impact the encoded data size. However, further work based on time windowing is necessary to compress vibrations with temporal variations.

The findings of this study hold significant value for a wide range of applications that involve delivering vibrations to users. Everyday human-computer interfaces, such as smartphones, tablets, and wearables, could greatly benefit from a sparse synthesizer capable of reproducing complex vibrations.
When combined with algorithms from existing literature to adapt to the user's finger movement in the case of active touch~\cite{friesen2023perceived, romano2010automatic}, the compression method could prove useful for replicating textures in virtual environments. 
Moreover, synthesis approximations might be negligible as long as the vibrations remain plausible in a given context and correspond to the user's expectations~\cite{rosenkranz2023perceptual}.
 
Vibration compression is also valuable for enhancing the musical experience through multichannel vibrations, especially for audiences with hearing impairments. The vibration analysis framework could be effectively integrated into sensors of robotic arms to interpret textures in a manner similar to humans, or to render the essence of tactile information to individuals with prosthetic hands.

\section{Conclusion}

In this paper, we investigated the perception of noisy, stationary vibrations. The stimuli were taken from a data base of friction-induced vibrations recorded during the exploration of textures at constant velocity.
We initially examined the perception of the intensity of these vibrations and developed a simple model based on the power within the frequency bandwidth of human optimal sensitivity. 
Subsequently, we investigated other perceptual attributes through dissimilarity experiments. These experiments revealed that the essential information regarding the spectral content resides in the balance between high and low frequencies.
The observed discrepancies were attributed to specific stimuli that were not entirely homogeneous, wherein pertinent information also lies in the phase. In summary, the perception of purely stationary noisy vibrations is driven by two main attributes: 1) intensity and 2) high/low frequency balance.
We showcased the potential of these findings for sparse analysis and synthesis of vibrations, as well as perception-based compression. Such applications can be valuable for any device delivering complex vibratory feedback.





\bibliographystyle{IEEEtran}
\bibliography{biblio}

\begin{IEEEbiography}[{\includegraphics[width=1in,height=1.25in,clip,keepaspectratio]{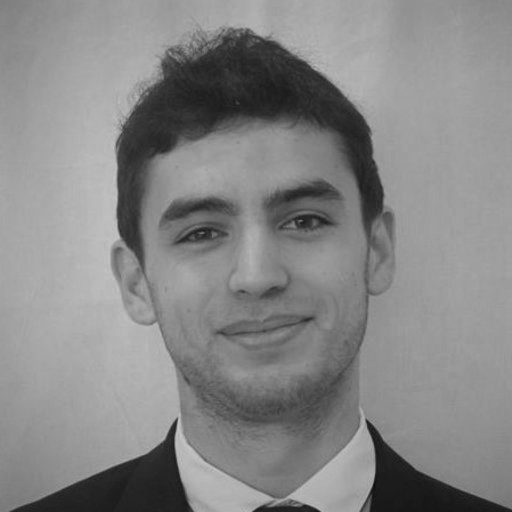}}]{Corentin Bernard}
 graduated from the Ecole Centrale de Marseille in 2017 and holds a Master’s degree in acoustics from the Aix-Marseille Universit\'e. In 2022, he earned his Ph.D. for research conducted at Stellantis and two laboratories: the Perception, Representations, Image, Sound and Music Laboratory (PRISM) and the Institute of Movement Sciences (ISM) in Marseille. He is now a post-doctoral researcher at the CNRS in the PRISM laboratory and the Mediterrean Institute of Robotics and Automation (MIRA).
 His research interests lie in understanding human perception of tactile stimulations and sounds. His work focuses on psychophysics, multisensory integration, and eyes-free human-machine intertaction.

\end{IEEEbiography}

\begin{IEEEbiography}[{\includegraphics[width=1in,height=1.25in,clip,keepaspectratio]{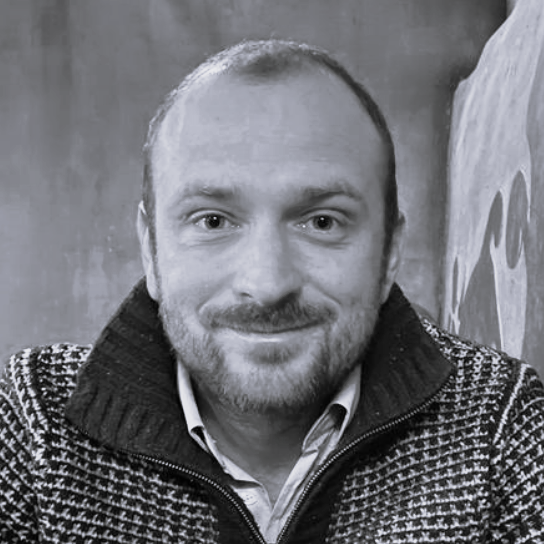}}]{Etienne Thoret} obtained a Ph.D. from Aix-Marseille University in 2014. He was then a post-doc at McGill University between 2015 and 2018 before a brief post-doctoral assignment at the Ecole Normale Supérieure de Paris. Later, he conducted research funded by the Institute of Language Communication and the Brain (ILCB) for four years, working at the crossroads of the Perception, Representations, Image, Sound, and Music (PRISM) lab and the Laboratoire d'Informatique et Systèmes (LIS) where he started to develop adaptive signal processing representations simulating cochlear processes fed to deep-neural networks to decipher auditory cerebral processes, the current topic of his research as a CNRS researcher (Centre National de la Recherche Scientifique) at the Institut de Neuroscience de la Timone (INT).
\end{IEEEbiography}

\begin{IEEEbiography}[{\includegraphics[width=1in,height=1.25in,clip,keepaspectratio]{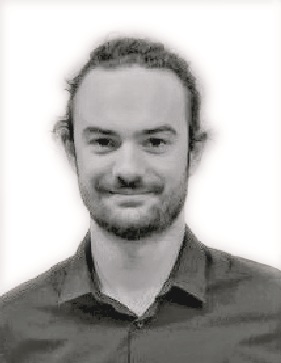}}]{Nicolas Huloux}
graduated from Grenoble-INP Engineering school in 2017 and obtained his PhD in Biorobotics from Aix-Marseille University in 2021 under the supervision of Michaël Wiertlewski. Following this, he initiated the development of a research robotics ecosystem in Ajaccio, concomitantly with the creation of the engineer school MIRA (Mediterrean Institute of Robotics and Automation). His research mainly focuses on multimodal human-machine interactions, with a specific emphasis on tactile perception at the interface. 
\end{IEEEbiography}

\begin{IEEEbiography}[{\includegraphics[width=1in,height=1.25in,clip,keepaspectratio]{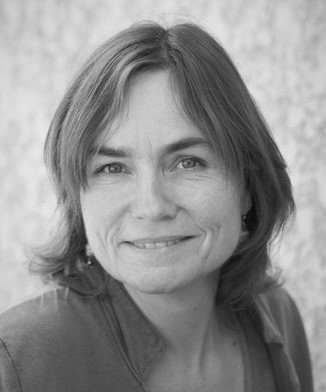}}] {S\o lvi~Ystad}
received her degree as a civil engineer in electronics from NTH (Norges Tekniske H\o gskole), Norway in 1992. In 1998 she received her Ph.D. degree in acoustics from the University of Aix-Marseille II, Marseille. After a post doctoral stay at the University of Stanford - CCRMA, California, she obtained a researcher position at the CNRS (Centre National de la Recherche Scientifique) in Marseille in 2002. In 2017 she co-founded the interdisciplinary art-science laboratory PRISM - Perception, Representations, Image, Sound, Music – (www.prism.cnrs.fr) in Marseille. Her research activities mainly focus on investigations of auditory perception through so-called perceptual engineering which consists of crossing different disciplines to link physical and signal knowledge with human perception and cognition. 
\end{IEEEbiography}

\end{document}